\documentclass[sigplan,screen,authorversion,nonacm]{acmart}

\usepackage{algorithm}
\usepackage[frozencache]{minted}
\usepackage{mathtools}

\newcommand{\pgraphsp}{\vspace{0.33em}}
\newcommand{\paragrapht}[1]{\pgraphsp \indent \textbf{\emph{#1}}}

\DeclarePairedDelimiter{\ceil}{\lceil}{\rceil}

\newcommand{\code}[1]{\mintinline[breaklines, breakafter=_]{C++}{#1}}
\newcommand{\shp}{\code{shp}}
\newcommand{\mhp}{\code{mhp}}
\newcommand{\domain}{memory locale}
\newcommand{\reserved}{\textsuperscript{®}}

\newcommand{\tile}{tile}

\newcommand{\nocomments}{}

\ifx\nocomments\undefined
\newcommand{\nbc}[3]{
	{\colorbox{#3}{\bfseries\sffamily\scriptsize\textcolor{white}{#1}}}
	{\textcolor{#3}{\sf\small$\blacktriangleright$\textit{#2}$\blacktriangleleft$}}
}
\else
\newcommand{\nbc}[3]{}
\fi

\definecolor{babcolor}{rgb}{0.9,0.45,0.1}
\definecolor{rccolor}{rgb}{0.06,0.07,0.62}
\definecolor{lscolor}{rgb}{0.62,0.15,0.06}
\definecolor{jkcolor}{rgb}{0.11,0.5,0.06}
\definecolor{tmcolor}{rgb}{0.5,0.1,0.6}

\newcommand{\bab}[1]{\nbc{BAB}{#1}{babcolor}}
\newcommand{\rc}[1]{\nbc{RC}{#1}{rccolor}}
\newcommand{\ls}[1]{\nbc{LS}{#1}{lscolor}}
\newcommand{\jk}[1]{\nbc{JK}{#1}{jkcolor}}
\newcommand{\tm}[1]{\nbc{TM}{#1}{tmcolor}}

\AtBeginDocument{%
  \providecommand\BibTeX{{%
    \normalfont B\kern-0.5em{\scshape i\kern-0.25em b}\kern-0.8em\TeX}}}

\begin{document}

\title{Distributed Ranges: A Model for Distributed Data Structures, Algorithms, and Views}

\author{Benjamin Brock}
\affiliation{%
  \institution{Intel Corporation}
  \city{Santa Clara}
  \state{CA}
  \country{USA}
}
\email{benjamin.brock@intel.com}

\author{Robert Cohn}
\affiliation{%
  \institution{Intel Corporation}
  \city{Salem}
  \state{NH}
  \country{USA}
}
\email{robert.s.cohn@intel.com}

\author{Suyash Bakshi}
\affiliation{%
  \institution{Intel Corporation}
  \city{Houston}
  \state{TX}
  \country{USA}
}
\email{suyash.bakshi@intel.com}

\author{Tuomas Karna}
\affiliation{%
  \institution{Intel Corporation}
  \city{Helsinki}
  \country{Finland}
}
\email{tuomas.karna@intel.com}

\author{Jeongnim Kim}
\affiliation{%
  \institution{Intel Corporation}
  \city{Hillsboro}
  \state{OR}
  \country{USA}
}
\email{jeongnim.kim@intel.com}

\author{Mateusz Nowak}
\affiliation{%
  \institution{Intel Corporation}
  \city{Gliwice}
  \country{Poland}
}
\email{mateusz.p.nowack@intel.com}

\author{Łukasz Ślusarczyk}
\affiliation{%
  \institution{Intel Corporation}
  \city{Radom}
  \country{Poland}
}
\email{lukasz.slusarczyk@intel.com}

\author{Kacper Stefanski}
\affiliation{%
  \institution{Intel Corporation}
  \city{Warsaw}
  \country{Poland}
}
\email{kacper.stefanski@intel.com}

\author{Timothy G. Mattson}
\affiliation{%
  \institution{Intel Corporation}
  \city{Ocean Park}
  \state{WA}
  \country{USA}
}
\email{timothy.g.mattson@intel.com}

\renewcommand{\shortauthors}{Brock and Cohn, et al.}

\begin{abstract}
Data structures and algorithms are essential building blocks for programs, and \emph{distributed data structures}, which automatically partition data across multiple \domain{}s, are essential to writing high-level parallel programs.
  While many projects have designed and implemented C++ distributed data structures and algorithms, there has not been widespread adoption of an interoperable model allowing algorithms and data structures from different libraries to work together.  This paper introduces distributed ranges, which is a model for building generic data structures, views, and algorithms.  A distributed range extends a C++ range, which is an iterable sequence of values, with a concept of segmentation, thus exposing how the distributed range is partitioned over multiple \domain{}s.  Distributed data structures provide this distributed range interface, which allows them to be used with a collection of generic algorithms implemented using the distributed range interface.
The modular nature of the model allows for the straightforward implementation of \textit{distributed views}, which are lightweight objects that provide a lazily evaluated view of another range.  Views can be composed together recursively and combined with algorithms to implement computational kernels using efficient, flexible, and high-level standard C++ primitives.
We evaluate the distributed ranges model by implementing a set of standard concepts and views as well as two execution runtimes, a multi-node, MPI-based runtime and a single-process, multi-GPU runtime.  We demonstrate that high-level algorithms implemented using generic, high-level distributed ranges can achieve performance competitive with highly-tuned, expert-written code.
\end{abstract}

\maketitle

\section{Introduction}

\bab{See lines 77-87 if you want a comment macro.}
\tm{Tim can leave comments like this.}
\rc{Robert can leave comments like this.}
\jk{Jeongnim can leave comments like this.}
\ls{Lukasz can leave comments like this.}


Data structures are an essential building block for productive programming,
allowing users to implement algorithms using high-level data structure
primitives such as insertion, deletion, and iteration instead of directly
manipulating raw data.
In distributed memory programming, a lack of widely available
distributed data structures has been cited as a major barrier to programmer
productivity~\cite{Fuerlinger:2016:DASH,Brock:2019:BCD:3337821.3337912}.
Many projects have developed distributed data structures over the past few
decades, although no cross-platform model for building and integrating distributed
data structures has gained acceptance.  Both language-based approaches~\cite{kennedy2011CACM,weiland2007chapel,yelick1998titanium,upc2005upc},
which provide more power in terms of static program optimization, and
library-based approaches~\cite{nieplocha1996global,chakrabarti1995multipol,Tanase:2011:SPC:1941553.1941586,Fuerlinger:2016:DASH,kaiser2020hpx,DBLP:conf/ccgrid/CastellanaM18, Brock:2019:BCD:3337821.3337912},
which can be easier to integrate with pre-existing codebases, have been developed.
Recent efforts have focused on providing parallel and high-performance versions
of standardized, high-level programming environments.
Many hardware vendors now provide
high-performance implementations of C++ standard library
algorithms~\cite{bell2012thrust,merrill2015cub,rocthrust2023,onedpl2023}.
These parallel versions of C++ standard library algorithms can
perform computation using multiple threads or by launching work on a
single GPU~\cite{bell2012thrust,merrill2015cub,rocthrust2023,onedpl2023}.
At the same time, these standard algorithms are flexible and generic and can
be used with many different types of data structures.  They do this by
depending on the iteration concepts defined in the standard library.  As long
as a data structure implements iterators that satisfy one of the standard
iterator concepts, they can be passed into one of these algorithms.  These
iterators are arranged in a hierarchy based on their capabilities, including
forward iterators, which can only iterate forward, as in a linked list;
bidirectional iterators, which can iterate both backwards and forwards, as in
a doubly linked list; random access iterators, which can iterate to any random
offset, as in an array; and contiguous iterators, which represent a contiguous
block of memory as in a C-style array.  A data structure implementer only has
to expose the generic iterator interface, and then their data structure can
plug into a collection of generic algorithms.

\begin{algorithm}[t]
\begin{minted}[fontsize=\small]{C++}
// using dr::shp or dr::mhp;

template <distributed_range X, distributed_range Y>
auto dot_product(X&& x, Y&& y) {
  auto z =   views::zip(x, y)
           | views::transform([](auto l, auto r) {
                                return l * r;
                              });

  return reduce(par_unseq, z.begin(), z.end());
}
\end{minted}
  \caption{A dot product implemented using standard C++ views and
           algorithms.  See Figure~\ref{fig:dot_product_diagram}.}
\label{alg:dot_product}
\end{algorithm}

The ranges library~\cite{cppranges} further generalizes the iterator interface
by introducing specific concepts for \emph{ranges}, which are sequences of
elements that can
be iterated over using \code{begin} and \code{end} iterators that they expose.
Most data structures are ranges, and some other objects, namely \emph{views},
are also ranges.  Views are lightweight objects that typically hold no data, but
fulfill the range interface, usually by providing a lazily evaluated,
modified view of another range. For example, a \code{transform_view} represents
a symbolic, lazily-evaluated view of another range with a binary function
applied to each element.  When views are combined with algorithms, they greatly expand
expressiveness, since they allow for transformations such as
combining multiple ranges, applying element-wise functions, or offsetting indices
to be fused together and optimized through inlining.
The combination of parallel algorithms with
high-level, composable views creates a high-level functional programming
environment where code written in standard C++ can automatically be executed in
parallel~\cite{haidl2017towards}.  Algorithm~\ref{alg:dot_product} demonstrates
a parallel dot product implemented using a combination of ranges, views, and
algorithms.  Today, such standard C++ programs can be executed across multiple
threads or on a single GPU using vendor-supplied algorithms.

For distributed programming environments, however, data is partitioned
over multiple \emph{\domain{}s}, such as GPUs within a single
server or nodes in a cluster.  In these situations, the standard range concept
is insufficient, since it does not expose any concept of segmentation.
While most distributed data structures libraries also have algorithms that can
operate on their data structures, there is no prevailing concept for
distributed data structures between libraries.  Most distributed data structures
have algorithms that operate only on a particular data structure, meaning that
for each new data structure a new implementation of each algorithm will be
required.  Without a generic distributed range concept, it quickly becomes
infeasible to write all the required algorithm implementations, and so
distributed data structures libraries seldom allow views that operate over
distributed data, and no distributed data structures libraries implement views
like those defined in the ranges library.

In this paper, we present the \emph{distributed ranges} model, which is a set
of concepts for operating over distributed data structures.  Algorithms can be
written generically using the distributed ranges concepts and later used with
any distributed data structure as long as it implements the distributed
range interface.
A distributed range is an iterable sequence of values, just like a standard range,
but it also exposes how its data is partitioned into multiple segments by
providing a \code{segments} customization point.  \code{segments} returns a
range containing all the segments, each of which is a \emph{remote range}, which
is a standard range with added locality information through a \code{rank}
customization point.

\begin{figure}[t]
  \centering
  \includegraphics[width=\columnwidth]{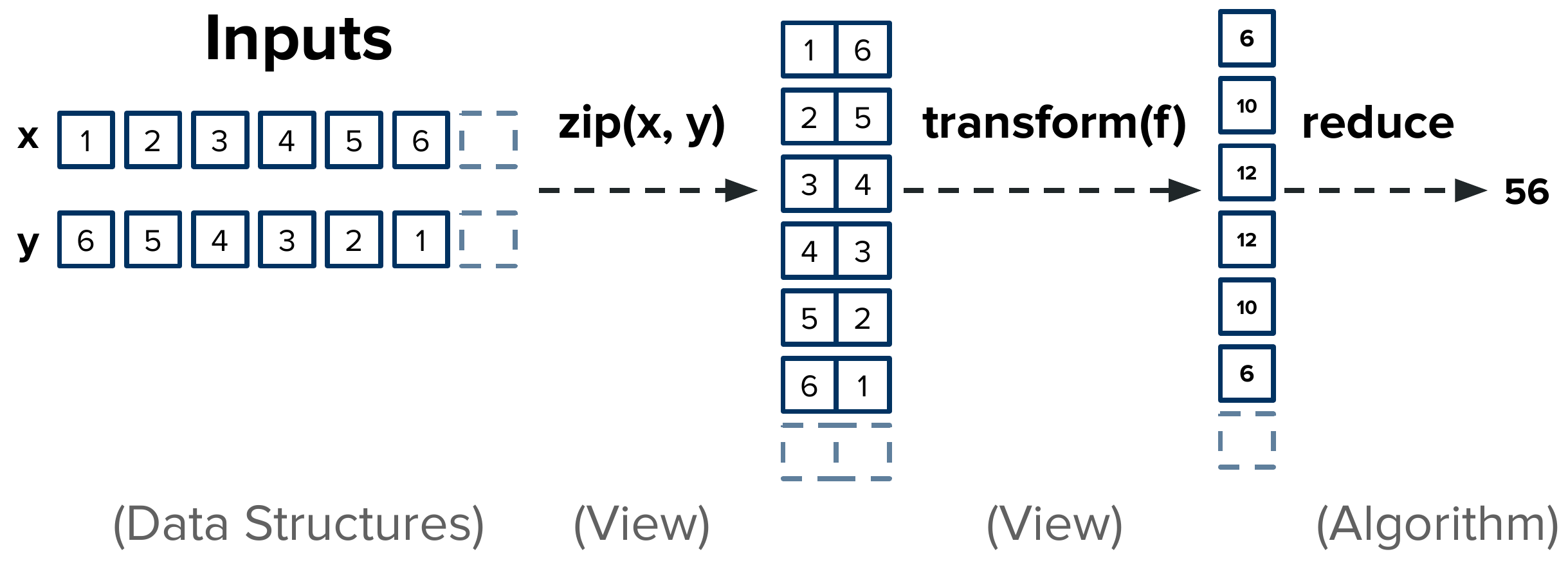}
  \caption{Diagram depicting the standard C++ dot product implementation in Algorithm~\ref{alg:dot_product}.
           First, corresponding elements of the two input ranges are
           zipped together to form a range of tuples.  Then, transform
           multiplies the two elements together.
           Finally, a parallel reduce is applied to
           compute the dot product in parallel. Transformations applied
           through views are lazy, allowing them to be fused
           and optimized through inlining.}
  \label{fig:dot_product_diagram}
\end{figure}

Being able to treat data structures generically, as long as they fulfill the
distributed range concept, greatly simplifies the implementation effort
necessary when adding a new data structure, since as soon as it fulfills
the distributed range interface it can be plugged in to a collection of
pre-existing algorithms.
In addition, the distributed ranges model enables the straightforward
implementation of lightweight, composable views that operate on distributed
ranges.  A view can simply take in another range as input, and, if that range is a
distributed range, fulfill the distributed range interface itself by exposing
an altered view of the base range's segments.  This composability allows views
to be stacked on top of each other, since a view can accept another view as
its base range so long as it exposes the distributed range interface.
The end result is that many standard C++ programs, such as the one shown in
Algorithm~\ref{alg:dot_product}, can be run across multiple GPUs or nodes with
little or no modification.

In this paper, we propose the distributed ranges model for generic distributed
data structures, algorithms, and views.  We present our implementation of
distributed ranges, which consists of a set of C++ concepts and customization
points, as well as a set of lightweight views that operate on distributed
ranges.  We also present two execution runtimes with distributed ranges algorithms
and data structures: (1) \shp{}, which supports single node, multi-GPU execution
using SYCL and (2) \mhp{}, which
supports multi-node execution on CPUs and GPUs using MPI one-sided
communication.
\\

This paper makes the following contributions:
\begin{enumerate}
  \item Propose the distributed ranges model, which generalizes the C++ standard
        library's range concept to support distributed data structures, algorithms,
        and views
  \item Implement two execution runtimes for distributed ranges, for single-process,
        multi-GPU execution and multi-process SPMD execution.
  \item Provide an implementation of cross-platform views that operate on distributed
        ranges.  This is the first distributed implementation of the C++ ranges
        library's views.
  \item Evaluate the performance of the distributed ranges abstraction on a set
        of standard C++ benchmarks, demonstrating that it can match the
        performance of vendor-optimized libraries on a single GPU and offer
        near perfect strong scaling up to 12 GPUs.
\end{enumerate}

\section{Background}
Distributed data structures provide a high-level abstraction to the user by
automatically partitioning data over multiple
\emph{\domain{}s}.  A \domain{} is a region of
memory that is remotely accessible to all processes, but is typically local to
some execution agent.  For example, in a multi-node program, remote direct
memory access (RDMA) can be used to make memory resident on
a particular process remotely accessible to the other processes.
However, the memory will still be fastest if used by the local process.
A distributed data structure will typically automatically partition data into
multiple \emph{segments}, each of which will be located in a particular \domain{}.
Users can use the distributed data structure as if it were a traditional data structure
using both data structure methods and algorithms, and the data structure will
ensure that the correct remote data is accessed.  In RDMA-based distributed
data structure libraries~\cite{Brock:2019:BCD:3337821.3337912,Fuerlinger:2016:DASH}
this is done by manipulating remote memory directly over the network.
In RPC-based distributed data structure
libraries this is done using remote procedure calls or active
messages~\cite{kaiser2020hpx,DBLP:conf/ccgrid/CastellanaM18,Tanase:2011:SPC:1941553.1941586}.

While many distributed data structures libraries implement distributed algorithms
that operate over data structures in
parallel~\cite{Brock:2019:BCD:3337821.3337912,Fuerlinger:2016:DASH},
these are typically data structure-specific algorithms, not algorithms that
can operate over any data structure.  For data structure libraries that provide
multiple different distributed data structures, this tends to limit their
utility, and there is no model that allows interoperability between data
structures and algorithms from different libraries.

C++20 introduced a number of language and library features that simplify the
creation of high-level libraries.  Here, we provide background on some of the
standard C++ features and concepts used in our work.

\paragrapht{Concepts}
C++ concepts describe a list of requirements for a type.  Similar to an interface
in other languages, a concept can require that a class supports a particular
method or that a particular expression involving a value of that type
be valid.  Concepts are particularly important for high-level libraries because
they allow methods to be written so that any type can be passed as an argument
to a function so long as it fulfills the requirements of one or more concepts.
This also greatly simplifies function overloading, as it allows different
overloads of a function to be written using different concepts, with the
appropriate function selected automatically at compile time.

\paragrapht{Customization Points and CPOs}
\rc{customization points are used in the implementation, but they seem like a detail that is much less important than the other subsections}
\emph{Customization points} are points where one library can customize its behavior when used
with another library.  For example, the method or function \code{begin} is a
customization point used to retrieve an iterator to the first element of a
range of values in C++.

Libraries can provide a customization point in multiple ways.  For
example, a user-defined array class could implement a \code{begin} method,
or it could implement a free-standing function \code{begin}.
As long as it returns an iterator, either one of these would be identified
by the ranges library as an implementation of the customization point \code{begin}
for the user-defined array type.  This is a key feature of customization
points, since implementing a customization point for a pre-existing class
does not necessarily require modifying it.

C++20 also introduced customization point objects (CPOs).  A CPO is a function
object that at compile time automatically selects the appropriate customization
point implementation.  For example, it may first check for a \code{begin}
method, and, if that does not exist, then look for a free-standing function
\code{begin} using argument-dependent lookup.

\paragrapht{Ranges Library}
Building on new mechanisms such as concepts and CPOs, C++20
introduced the \textit{ranges library}~\cite{cppranges}.  Ranges build on top of iterators,
which are generalized pointers to the beginning and end of a sequence
of values.  A range is a type that implements the \code{begin} and
\code{end} customization points, exposing those iterators.  Having objects that
represent an entire range instead of individual iterators greatly simplifies
the implementation of \emph{views} and \emph{algorithms}.
Views are lightweight objects that usually provide a non-owning view of another
range.
Commonly used views include \code{transform_view},
which applies an element-wise unary function similar to a \emph{map} operation,
\code{zip_view}, which zips together corresponding elements of two or more
ranges, and \code{take_view}, which creates a view of the first $n$ elements
of a range.  Views can be used recursively, effectively fusing multiple
transformations together at compile-time. Views enable the high-level
implementation of parallel kernels when used together with standard parallel
algorithms.

\begin{table*}[t]
    \centering
    \caption{Distributed ranges concepts, along with the customization points
             a type must implement to fulfill a concept.}
    \resizebox{\linewidth}{!}
    {
      \begin{tabular}{p{50mm} l r p{72mm}}
        \toprule
        Concept & Subsumes & Customization Points & Comment\\
        \toprule
          \code{remote_range} & \code{forward_range} & \code{rank}, *\code{local}
             & A \code{remote_range} is a \code{forward_range} with a \code{rank} stating which \domain{} it is in.\\

        \code{contiguous_remote_range} & \code{remote_range} & \code{rank}, \code{local}
             & A \code{contiguous_remote_range} is a \code{remote_range} with a \code{local} that returns a \code{contiguous_range} valid to use in the \code{rank} execution agent.\\

        \code{distributed_range} & \code{forward_range} & \code{segments}
             & A \code{distributed_range} is a \code{forward_range} with a \code{segments} that returns a range of \code{remote_range}s.\\

        \code{contiguous_distributed_range}
        \phantom{jkl jkl jkl jkl jkl jkl jkl jkl jkl jkl jkl jkl jkl} *optional
        & \code{distributed_range} & \code{segments}
             & A \code{contiguous_distributed_range} is a \code{distributed_range} with a \code{segments} that returns a range of \code{contiguous_remote_range}s.\\
        \bottomrule
        \end{tabular}
    }
    \label{table:concepts}
\end{table*}

\paragrapht{Parallel Algorithms}
C++17 introduced parallel versions of algorithms such as \code{reduce},
\code{inclusive_scan}, and \code{sort}.  The C++ standard supports
parallel execution using generic execution policies such as \code{par_unseq},
which allows parallelized and vectorized execution.
Standard library implementations GNU libstdc++ and libcxx use these standard
execution policies to support multi-threading and vector parallelism.  In
addition, many hardware vendors provide their own implementations of C++
standard algorithms, including Intel's oneDPL~\cite{onedpl2023},
Nvidia's Thrust~\cite{bell2012thrust} and CUB~\cite{merrill2015cub}, and
AMD's rocThrust~\cite{rocthrust2023}.
These implementations provide not only multi-threaded CPU implementations
but also launching work on GPU or FPGA devices using implementation-defined
execution policies.

\section{Distributed Ranges}
In order to implement generic distributed data structures
that can be used together with algorithms and views, we propose a generalization
of the ranges library called distributed ranges.  This model is built
on the concept \code{distributed_range}\footnote{For clarity, we use
monospace script (\code{distributed_range}) for the concept and
regular script (distributed range) for an object that fulfills the concept.},
which is analogous to the standard
\code{range}, but can be split up over multiple \domain{}s.
There are three requirements for such a concept:

\begin{enumerate}
  \item Algorithms can be written \emph{generically} using the\\ \code{distributed_range}
        concept and then used with all distributed data structures.
        (Although specializations can be written for improved performance.)
  \item Views can be created that take in a distributed range and then expose
        the \code{distributed_range} concept themselves.  Views can be nested recursively.
  \item The \code{distributed_range} concept must expose the overall distribution of
        the range, as locality information is essential for performance.
\end{enumerate}

A distributed data structure commonly consists of multiple segments, where each
segment may be located in a different \domain{}.  Together, these segments
make up the complete data structure.  Our model begins by adding a concept of
locality using a concept called a \code{remote_range}.


\paragrapht{Remote Range}
To model a segment that lives in one particular \domain{}, we define a
concept called a remote range.  A \code{remote_range} is a
standard C++ \code{forward_range}, with the
refinement that it also implements a customization point called \code{rank}, which
provides information about where the range is located.  A \code{remote_range} is
accessible from all execution agents in the program, but since it is located in
a particular \domain{}, it will be most efficient to access it from
an execution agent associated with that \domain{}.  For communication frameworks
in which memory is not directly accessible using raw language pointers from all
execution agents, such as RDMA-based communication frameworks, a \code{remote_range}
must have a remote iterator type that automatically triggers calls to the
communication library in order to access elements of the range.  This can be
performed using the normal standard C++ mechanisms of \code{copy} to copy
data into and out of the range as well as the dereference \code{operator*}.

In the case that these remote iterators add additional overhead when accessing
local data, as is commonly the case with distributed memory frameworks, the
customization point \code{local} may be provided, returning a local view of
the remote range that is valid only on the execution agent specified by
\code{rank}.
\rc{a figure with distributed range and remote range would be appropriate.}

To implement the \code{rank} customization point, a remote range can
implement a \code{rank} method \textit{or} a free-standing \code{rank} function that
accepts the range as a parameter.  As long as one of these exists and returns
an integer value, it will be identified by the
\code{rank} CPO, thus fulfilling the \code{remote_range} concept as defined in
Algorithm~\ref{alg:range_concepts}.  Being able to implement the customization
point as a free-standing function is important, since it enables us to turn
pre-existing data structures and views into remote ranges without modifying them.  We can
do this by implementing the customization point as a free-standing function.


\paragrapht{Distributed Range}
While a remote range represents a range that is located in a single \domain{},
a distributed range may be split up over more than one \domain{}.
A \code{distributed_range} is a standard C++ \code{forward_range} that also implements
the customization point \code{segments}.  \code{segments} returns a range of
\code{remote_range}s, where each \code{remote_range} corresponds to one segment of the
\code{distributed_range}.
Concatenated together, these segments form the complete logical \code{distributed_range}.

Since a \code{distributed_range} is a \code{forward_range}, but also has a \code{segments}
customization point, there are two ways to access the elements of a
distributed range.  The first is to iterate through the range as
a normal range.  While this may be inefficient, it is useful to have this
capability for printing and debugging.  The second mechanism for
accessing the range, the \code{segments} customization point, exposes
the distribution of the \code{distributed_range}, including how many segments the
range has, where each segment is located, and how large each segment is.  This
distribution information allows algorithms to intelligently place work on
different execution agents for processing, or even to apply load balancing or
work stealing algorithms based on the distribution of data within a range.
A \code{distributed_range} can have a wide variety of different distributions, since
each segment may have a different size and \domain{}.
The \code{segments} customization point
can return a view, which means that the object returned from \code{segments},
corresponding to the logical layout of the array, does not necessarily have to
correspond to the physical layout of the array.  For example, for a block cyclic
array, the data structure could store all blocks on each process contiguously,
returning a lazily evaluated view of the logical segments from the \code{segments}
customization point.

To implement the \code{segments}
customization point, a range can have a \code{segments} method or a free-standing
\code{segments} function that returns a range of remote ranges.  As long as one of
these implementations is available, it will be identified by the \code{segments}
CPO, thus fulfilling the \code{distributed_range} concept as defined in
Algorithm~\ref{alg:range_concepts}.  The ability to implement customization
points as a free-standing function is again important, as it enables us to
promote pre-existing classes, such as views, to distributed ranges without
modification.

\begin{algorithm}[t]
\begin{minted}[fontsize=\small]{C++}
template <typename T>
concept remote_range = forward_range<T> &&
                      requires(T& t) { rank(t); };

template <typename T>
concept distributed_range = forward_range<T> &&
                      requires(T& t) { segments(t); };
\end{minted}
  \caption{Concepts for remote and distributed ranges.  A \code{remote_range} is a
           standard \code{forward_range} that also has the \code{rank} customization point,
           indicating locality.  A \code{distributed_range} is a standard \code{forward_range}
           that also has the \code{segments} customization point, returning the distributed
           range's segments.}
\label{alg:range_concepts}
\end{algorithm}

\paragrapht{Additional Concepts}
In addition to these fundamental concepts, we also implement some refinements
of remote and distributed ranges similar to those in the standard library.
For example, a \code{remote_contiguous_range} is a \code{remote_range} backed by a block of
contiguous memory, meaning its contents can be copied using a single
\code{memcpy} operation.  Similarly, a \code{distributed_contiguous_range} is a
distributed range whose segments are all \code{remote_contiguous_range}s, meaning that
it is a distributed array whose segments are all contiguous blocks of remote
memory.  Additional concepts such as these, which refine the original distributed
and remote range concepts, can allow for optimizations inside of algorithms.


\section{Implementation}
Our prototype implementation of the distributed ranges has three modular
components:
\begin{enumerate}
  \item A collection of concepts and CPOs supporting distributed ranges, as well as
implementations of standard C++ views supporting distributed ranges.
  \item \shp{}, a single-process execution runtime along with data structures,
  algorithms, and communication primitives implemented in
  SYCL~\cite{keryell2015khronos}, supporting multi- CPU, GPU, and FPGA
  execution within a single node.
  \item \mhp{}, a multi-process execution runtime along with data structures,
  algorithms, and communication primitives implemented using one-sided
  MPI~\cite{gerstenberger2014enabling}.
\end{enumerate}

Separate execution frameworks are required because single- and multi-process
execution require separate execution models.
Like most distributed memory programs, \mhp{} uses a SPMD execution model in which
the same program is executed on multiple processes.  Operations such as constructing
a new distributed data structure or invoking an algorithm must be called
collectively by all processes, while operations such as accessing a particular
segment of a data structure can be performed by a single process.

\shp{} uses a single-process execution model.  To perform work in parallel,
it asynchronously launches work on multiple SYCL devices.  In practice, programs
written using standard C++ data structures, algorithms, and views can
be executed using both \shp{} and \mhp{}, as shown in Algorithm~\ref{alg:dot_product}.


\subsection{Communication Operations}
\hspace{0em}\paragrapht{Remote Pointers}
Both \shp{} and \mhp{} expose a partitioned global address space (PGAS) model, in
which memory is globally accessible, but partitioned over multiple \domain{}s.
This is exposed through remote pointer types, which are smart pointers
that behave similarly to regular language pointers, but may reference memory
in another \domain{}.  When remote pointers are acessed, this triggers a
library call to read or write to remote memory.  In \shp{}, this is done using
SYCL memcpy commands when on the host or raw pointer operations on the device.
In \mhp{}, this is done using \code{MPI_Put} and \code{MPI_Get} operations.

Remote pointers are iterators and support dereferencing using \code{operator*},
which returns a remote reference object that serves as a proxy reference to the
underlying remote memory.  The standard \code{copy} and \code{memcpy} can also be 
used for copying data into, out of, and between remote memory buffers.

\paragrapht{Asynchronous Communication}
\shp{} uses SYCL, whose built-in asynchronous memcpy operation are used to
implement synchronous and asynchronous versions of \code{copy}.
\mhp{} uses the relaxed memory model provided by MPI RMA, in which writes return
without blocking while reads block until the data is ready to be read.
To ensure consistency, collective operations perform a barrier/memory flush
to ensure all writes are complete before returning control.
Collective operations employ asynchronous communication internally.

\subsection{Data Structures}
\label{sec:data_structures}
In our proof of concept implementations, we implemented several data structures
fulfilling the distributed ranges concept.  These data structure automatically
distributed data over multiple \domain{}s: over multiple GPUs in the case
of \shp{} and over multiple processes in the case of \mhp{}.  Each data structure
will automatically allocate multiple segments located in each \domain{}. 
Upon accessing data using a global indexing mechanism, such as \code{operator[]},
the data structure will automatically index into the correct segment, returning
a remote reference to the underlying data.

In addition, each data structure implements a \code{segments} method, which
returns a view of the segments that make up the distributed data structure.
Each element of the segments view is a \code{remote_range} representing a
segment of the overall data structure that is located in a particular
\domain{}.
A data structure's segments may directly correspond to the underlying
memory---such as in our distributed vector, where the segments model a remote
contiguous range---or, they may represent a more
complex derived view of the underlying data.  For example, in our distributed
matrix data structure, each segment is a \code{dense_matrix_view}, which,
encapsulating the 2D structure of the matrix, is more complex than a simple
contiguous range while still satisfying the \code{remote_range} concept.

\subsubsection{Distributed Vector}
Our distributed vector data structure provides a one-dimensional array that is
automatically distributed over multiple \domain{}s.  To construct a
\code{distributed_vector} of size $n$
distributed over $p$ \domain{}s, a block partitioning is used.
$p$ segments are allocated with room for $\ceil*{n/p}$ elements each.
In order to fulfill the \code{distributed_range} concept, \code{distributed_vector}
has a \code{segments} method that returns a view holding all of these
segments.  Since each segment is backed by a single block of contiguous memory,
each segment satisfies the concept \code{remote_contiguous_range}, and therefore the whole
\code{distributed_vector} also models \code{distributed_contiguous_range}.

Our distributed vector supports indexing with
\code{operator[]}, which
will internally perform index arithmetic to identify which index in which
segment needs to be accessed.  The \code{begin} and \code{end} methods provide
iterators over the global data structure.  Internally, these iterators will
scan over each segment of the distributed vector to traverse the entire global
vector.  These iterators satisfy the \code{distributed_iterator} concept, so
they can be passed into iterator-based algorithms.  Finally, the \code{segments}
method allows direct user access to each distributed vector segment
as well as allows the distributed vector to be passed into algorithms that
operate on distributed ranges.

\subsubsection{Distributed Matrix}
We implement two distributed matrix data structures, a distributed dense matrix 
data structure, which automatically distributes a dense matrix among multiple
\domain{}s, and a distributed sparse matrix data structure, which
automatically distributes a block CSR sparse matrix.  Both the dense and sparse
versions utilize the same user-configurable block partitioning strategy.

\paragrapht{Matrix Access}
Elements of the dense matrix can be accessed globally using \code{operator[]},
which returns a remote reference to the corresponding element.  Both the dense
and sparse matrix data structures also support an iterator interface based on
the GraphBLAS C++ API~\cite{brock2022grb_iterators},
where the iterator iterates over every stored element of the matrix as a tuple
value holding the corresponding row and column index as well as scalar value.

\paragrapht{Tile Partitioning}
Our distributed matrix data structure automatically partitions a matrix into
multiple tiles using a block cyclic distribution strategy that is user
customizable.
Users can provide explicit parameters to the distributed matrix in the form of
tile dimensions, which split the matrix into a tile grid, and a processor grid,
which assigns each tile to a \domain{},
or they can request a distribution using high-level block partitioning
descriptors such as \code{block_cyclic}, \code{block_row}, and \code{block_column}.
Each tile is a \code{remote_range} matrix view that can be viewed individually using the
\code{tile} method.  Local copies of tiles can also be retrieved using
\code{get_tile} and \code{get_tile_async} methods.

\subsection{Algorithms}
The distributed ranges concepts allow algorithms to be implemented in a
hierarchical fashion by decomposing the distributed range into its segments
and applying local versions of algorithms to each segment in parallel.
Crucially, distributed algorithms can be written generically, using only the
distributed range concept, and then applied to any data structure that satisfies
the distributed range concept.

\paragrapht{\code{for_each}}
The \code{for_each} algorithm, which executes a function for each
element of an input range, is straightforward to implement using distributed
ranges.  The distributed \code{for_each} algorithm iterates over each segment
in a distributed range, then launches a local version of \code{for_each} on
each segment using the execution agent associated with that segment's rank.
The \shp{} implementation asynchronously launches kernels on the devices
associated with each segment's \code{rank}, then waits for them to complete.
The \mhp{} implementation uses SPMD semantics to launch a local \code{for_each}
on each segments' corresponding process.

\paragrapht{\code{reduce}}
The \code{reduce} algorithm also works by launching a local algorithm on each
segment, as shown in Algorithm~\ref{alg:reduce}.  The local algorithms produce
partial reductions, which are then combined together to produce the final
result
\footnote{Our reduce and scan algorithms work with any commutative binary
operator, despite the simplification to standard arithmetic in
Algorithms~\ref{alg:reduce}-\ref{alg:inclusive_scan}.}.
In \shp{}, we iterate through each segment, launching a \code{reduce_async}
algorithm that returns a future that will contain the local result.  Once all the
local algorithms are launched, we wait for the kernels to complete and add the
results together.  In \mhp{}, each process launches a local \code{reduce} on
each of its segments, followed by an \code{MPI_Reduce} to produce the final
result.

\begin{algorithm}[t]
\begin{minted}[fontsize=\small]{C++}
template <distributed_range R, typename T>
void reduce(R&& r, T init) {
  // Iterate through all segments in parallel.
  parallel_for(auto segment : segments(r)) {
    // Get memory locale associated with
    auto rank = rank(segment);

    // Get execution policy associated with rank,
    // launch local reduce algorithm.
    auto policy = get_policy(rank);
    init += reduce(policy,
                   begin(segment), end(segment));
  }
  return init;
}
\end{minted}
  \caption{Pseudocode for \code{reduce} algorithm.}
\label{alg:reduce}
\end{algorithm}

\paragrapht{\code{inclusive_scan}}
\label{sec:inclusive_scan}
Our \code{inclusive_scan} algorithm is implemented by first performing a local
scan on each segment, with the result written into a temporary array \code{temp}.
After these scans are complete, the rightmost element of each array
\code{temp[temp.size()-1]} will contain the sum of all elements in the segment.
These elements are written into a local array on rank 0, and a scan is performed
on these elements.  Element $i-1$ of this array now holds the sum of all elements
before segment $i$ in the original range.  The corresponding element is added
to each to each segment to produce the final result.  This is then copied from a
temporary buffer into the output buffer.

In both \mhp{} and \shp{}, the temporary buffer can be avoided and the result
written directly into the user-provided output output buffer if the input and
output ranges \textit{align}, meaning that they have the same number of segments
and corresponding segments all have the same sizes and ranks.  In \shp{}, the
temporary buffer is avoided even for non-aligned ranges by using a \code{zip}
view, which produces a new set of overlapping segments as discussed in
Section~\ref{sec:zip_view}.  We currently prevent \code{zip} views with
non-aligned segments from being created in \mhp{}, since processing them naively
would be prohibitively inefficient.  Other than these differences, the
implementations in \shp{} and \mhp{} are the same, with work being launched using
SPMD-style execution or asynchronous kernel launches respectively.
A similar algorithm, with minor modification, is used to implement
\code{exclusive_scan}.

\rc{mhp implementation is different, need to understand why}

\begin{algorithm}
\begin{minted}[fontsize=\small]{C++}
template <distributed_range R, distributed_range O>
void inclusive_scan(R&& r, O&& o) {
  using T = range_value_t<R>;
  local_vector<T> partial_sums(zip(r, o).size());

  int segment_id = 0;
  parallel_for(auto [in, out] : segments(zip(r, o))) {
    // Call local inclusive_scan
    auto policy = /* ... */;
    inclusive_scan(policy, in, out);

    // Add the final element to the partial results.
    partial_sums[segment_id++] = *(out.end() - 1);
  }

  // Perform scan over partial sums to compute offsets.
  inclusive_scan(partial_sums, partial_sums);

  // Add correct offset to each segment.
  parallel_for(auto&& [in, out] : zip(r, o)) {
    sum = partial_sums[segment_id-1]
    for_each(out, [](auto&& v) {
                    v += sum;
                  });
  }
}
\end{minted}
\caption{Pseudocode for \code{inclusive_scan} algorithm.}
\label{alg:inclusive_scan}
\end{algorithm}

\paragrapht{\code{sort}}
We implement sort using a distributed partition sort algorithm.
First, each segment is sorted in place by its corresponding execution agent.
Next, $n - 1$ medians are selected for each segment to partition
the data into $n$ chunks where $n$ is the number of segments in the output
range.  These medians are then copied into a temporary buffer on rank 0,
where they are again sorted and the final $n - 1$ medians are selected to be used as
splitters for partitioning the data.  Given these medians, each agent counts how
many of its elements will be sent to each chunk, and these values are reduced
globally to compute the size of each chunk.  After the chunks are allocated, the
data is redistributed to the temporary chunks, then copied into the input range
\footnote{Note that C++'s \code{sort} is in place.}.
Each execution agent then copies the data from its segments into the corresponding
output buffer based on the splitters.  Finally, each execution agent sorts its
local buffer, and this sorted output is copied into the output range.
The implementation is similar in \shp{} and \mhp{}, with the exception that \shp{}
uses shared memory and queues for the histogram and redistribution stages, while
\mhp{} uses \code{MPI_Reduce} and \code{MPI_Alltoallv}.


\subsection{Views}
Views are lightweight objects that usually provide a non-owning,
lazily-evaluated view of another range.
They often perform lazy transformations to data, such as applying an elementwise
unary function, modfiying the size of a range, or combining multiple ranges
together.  Views provide the same range interface as data structures, allowing
them to be iterated over and used in algorithms just like normal data
structures.  In distributed ranges, we provide a set of views that operate on
normal ranges, but also on remote and distributed ranges by offering the
\code{rank} and \code{segments} CPOs if they are provided by the underlying
base range to which the view is applied.  Distributed and remote versions of
most views are implemented simply by taking a pre-existing implementation,
then implementing \code{rank} and \code{segments} customization points for it
as standalone functions.

To support the \code{rank} CPO, all views simply need to check whether the
base range used to create the view is a remote range.  If this is the case,
the \code{rank} CPO is implemented by returning the value of \code{rank} when
invoked on the base range.  The \code{segments} CPO required by
\code{distributed_range}, however, requires a different implementation for each
view type.

\paragrapht{Transform}
The transform view applies an elementwise function to each value of a range,
analogous to a map operation in functional programming environments.
To implement the \code{segments} CPO for transform views whose base is a
distributed range, we can return a view of the
base range's segments in which each segment has a transform view applied to it
using the same elementwise function.  Since each segment is a remote range,
the transform view of each segment will automatically be a remote range as
well.

\paragrapht{Take and Drop}
The take view shortens the length of a range to take at most the first $l$
elements of the range.  Similarly, the drop view shortens an array by dropping
the first $f$ elements.  To implement take and drop, we implement a helper
function called \code{trim_segments}.  \code{trim_segments} takes in a range of
segments as well as a two indices, $f$ and $l$, referring to positions in the
underlying elements.  \code{trim_segments} then returns a ``trimmed'' view of these
segments such that only the $f$'th through $l$'th elements are included.


\paragrapht{Zip}
\label{sec:zip_view}
The zip view takes in two or more ranges and returns a view of those ranges with
the corresponding elements \emph{zipped} together as a tuple.  Zips are useful
for performing operations that combine multiple arrays, such as a dot product
as demonstrated in Algorithm~\ref{alg:dot_product}.

Unlike the previously discussed views, a zip view involves multiple ranges.
When operating on two or more distributed ranges, the inputs may or may not be
\textit{aligned}, meaning they have the same number of segments and each set of
corresponding segments has the same size and rank.  In \mhp{}, naively operating
on mixed local and remote data would incur considerable performance overhead,
since it would require fine-grained remote reads and writes over the network as
elements are read from and written to.  In \shp{}, however, execution takes place
within a single node, and we are guaranteed to have SYCL peer-to-peer access,
meaning each device supports load-store semantics to other devices' memory.
This load-store communication is typically performed over a high-performance
fabric such as Xe Link~\cite{gomes2022ponte}, which has considerably lower latency and higher
bandwidth than going out over the network.  This means that naively reading and
writing individual elements will likely achieve good performance in \shp{},
whereas \mhp{} will require bulk network operations to achieve good performance.
For this reason, we prohibit the creation of zip views with non-aligned ranges
in \mhp{} while allowing them to be used in \shp{}, creating sets of contiguous
overlapping segments as described below.

For zip views where one or more of the base ranges is a remote range and no
range is a distributed range, we provide a \code{rank} method.  By convention,
we select the first remote range among the base ranges and return its rank.
When there are multiple remote ranges stored in different \domain{}s,
communication will be necessary in order to access the remote segments.  This
communication overhead will be particularly burdensome in \mhp{} due to the
need to access individual elements as discussed above, so this will result in
an error in \mhp{}.


When zipping together two
or more distributed ranges that have identical segments distributions, which is
the common case, we can implement the \code{segments} method for zip by
returning a view containing each of the corresponding segments zipped together.
If the ranges are \textit{aligned}, meaning they have the same number of segments
and the segments are the same size and rank, this is straightforward.
However, in the case that the base distributed ranges have nonaligned segments,
we must further partition the segments of one or both arrays to produce a set of
aligned segments.  We can do this by iterating through all of the segments using
two sets of iterators.  First, we maintain a tuple, \code{seg}, holding iterators
to the segments for each range in the zip.
Second, we maintain a tuple of iterators \code{local}
pointing to the elements currently being extracted from each segment.  At
initialization, \code{seg} holds iterators to the first segment in each
base range, and \code{local} holds iterators to the first element in each
segment in \code{seg}.
While there is still no iterator in \code{seg} that has not reached the end
of its corresponding segments range, we identify the largest segment that can be
created by computing the minimum distance between each element of \code{local}
and the corresponding sentinel \code{end} iterator for the segment, obtained
using \code{seg}.  We then use \code{take} to create a view of the corresponding
parts of each segment before zipping them together.  Finally, we increment each
iterator in \code{local}, incrementing the corresponding \code{seg} and moving on
to the next segment if the end of the segment has been reached.  Finally, once
one of the segments ranges ends, we return the newly created zipped segments.


\section{Evaluation}
To evaluate the performance and usability of the distributed ranges model, we
ran experiments on a multi-GPU system.
First, we ran a set of standard C++ performance benchmarks that can be
executed in parallel using distributed ranges with little or no modification.
We compare the performance of these benchmarks to single-GPU versions implemented
using vendor-provided versions of standard algorithms.
In addition, we evaluate the scalability of a distributed linear algebra benchmark
implemented in \shp{} using high-level distributed matrix data structures that
expose the distributed ranges model.

\subsubsection*{Evaluation System}
We ran experiments on a
system equipped with 6 Intel\reserved{} Data Center GPU Max 1550 GPUs, codename
Ponte Vecchio (PVC).  Each PVC GPU is split into two \tile{}s, each tiling having
64 GB of HBM2e memory.  The \tile{}s on a GPU can access each others' memory
using an inter-\tile{} interconnect providing 230 GB/s of
unidirectional link bandwidth.  All twelve \tile{}s can also access each
others' memory using an intra-node Xe Link interconnect offering 20 GB/s of
unidirectional link bandwidth per tile.  Our system is equipped with two
Intel\reserved{} Xeon\reserved{} CPU Max 9470 CPUs and 1 TB of memory.
All code was compiled using the Intel\reserved{} oneAPI C++ compiler version 2023.1.0.

\subsection{Standard C++ Benchmarks}
The distributed ranges model allows users to program using standard C++
views and algorithms and to execute that code across multiple GPUs or
nodes in a cluster.  The standard C++ benchmarks evaluated here require
minimal modification to run using distributed ranges.  The algorithms themselves
are unchanged except for including the \code{dr::shp/mhp} namespace instead of
\code{std::ranges} and using the \code{distributed_range} concept in the function
definition so that the function will throw an error unless called with a distributed
range.  The harness code must also use distributed data structures
as described in Section~\ref{sec:data_structures} when allocating data.
Algorithm~\ref{alg:dot_product} shows an implementation of the
dot product algorithm using standard C++ views and algorithms as provided by
distributed ranges.

\begin{figure*}
  \includegraphics[width=0.32\textwidth]{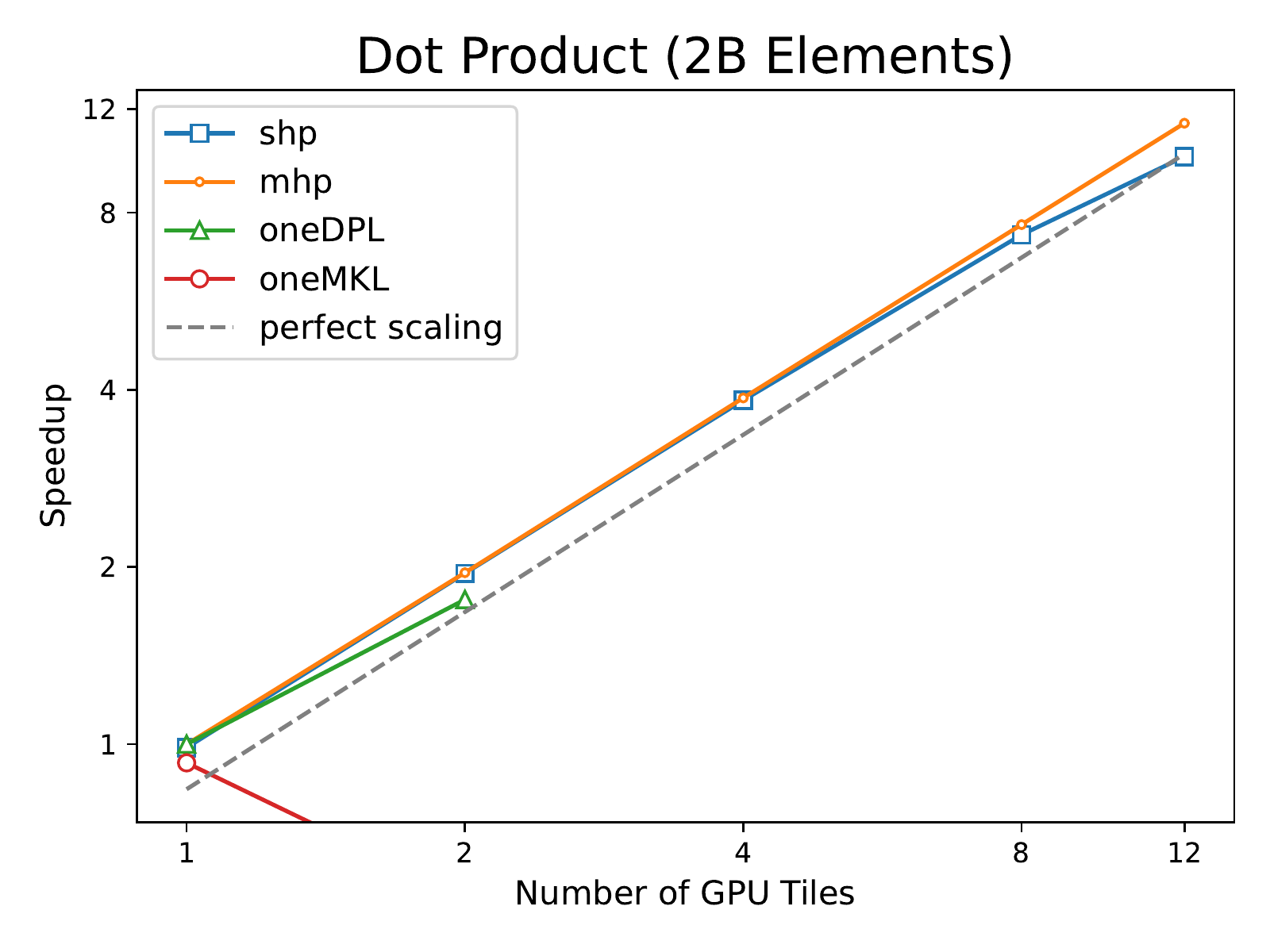}
  \includegraphics[width=0.32\textwidth]{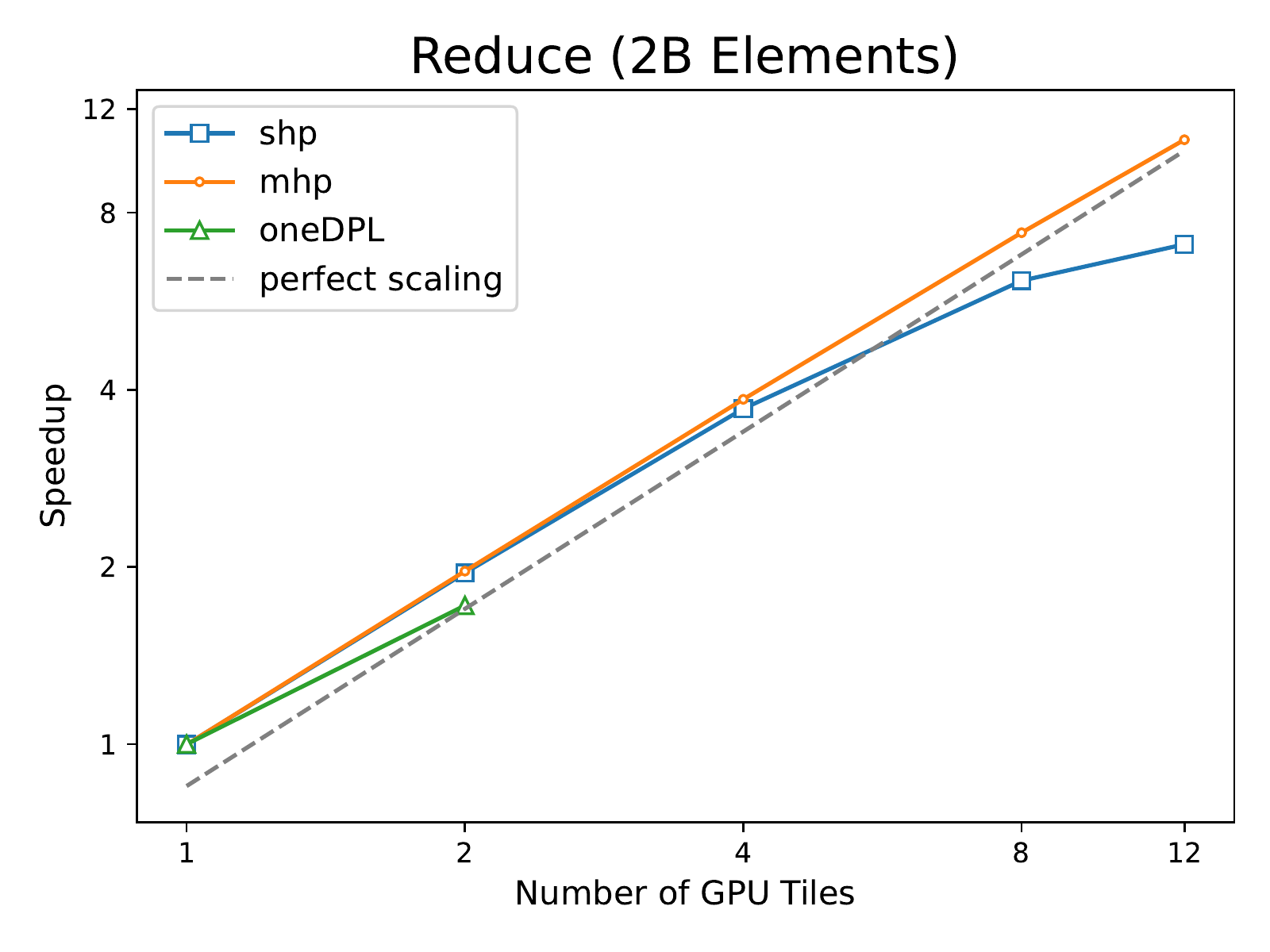}
  \includegraphics[width=0.32\textwidth]{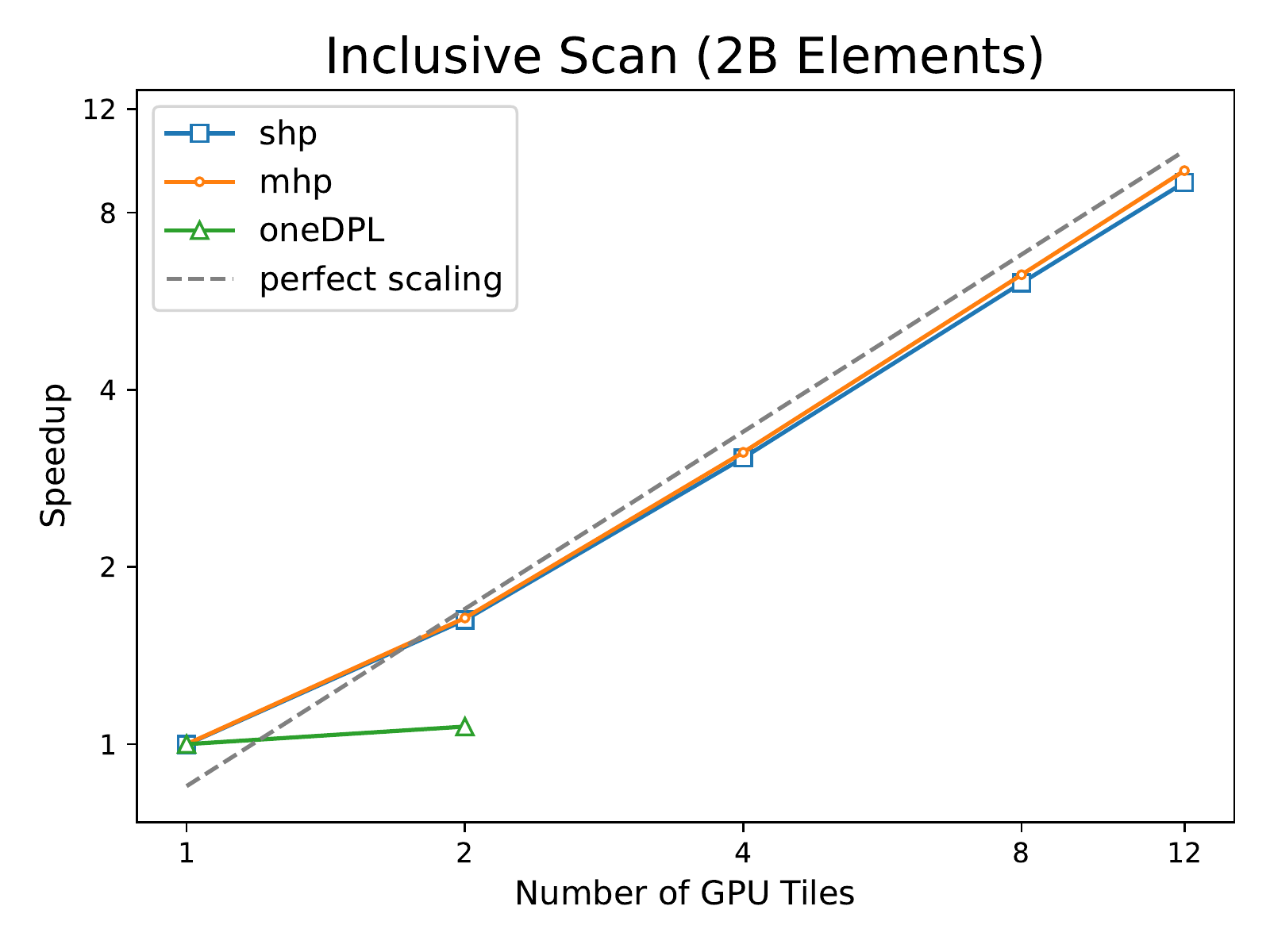}\\
  \includegraphics[width=0.32\textwidth]{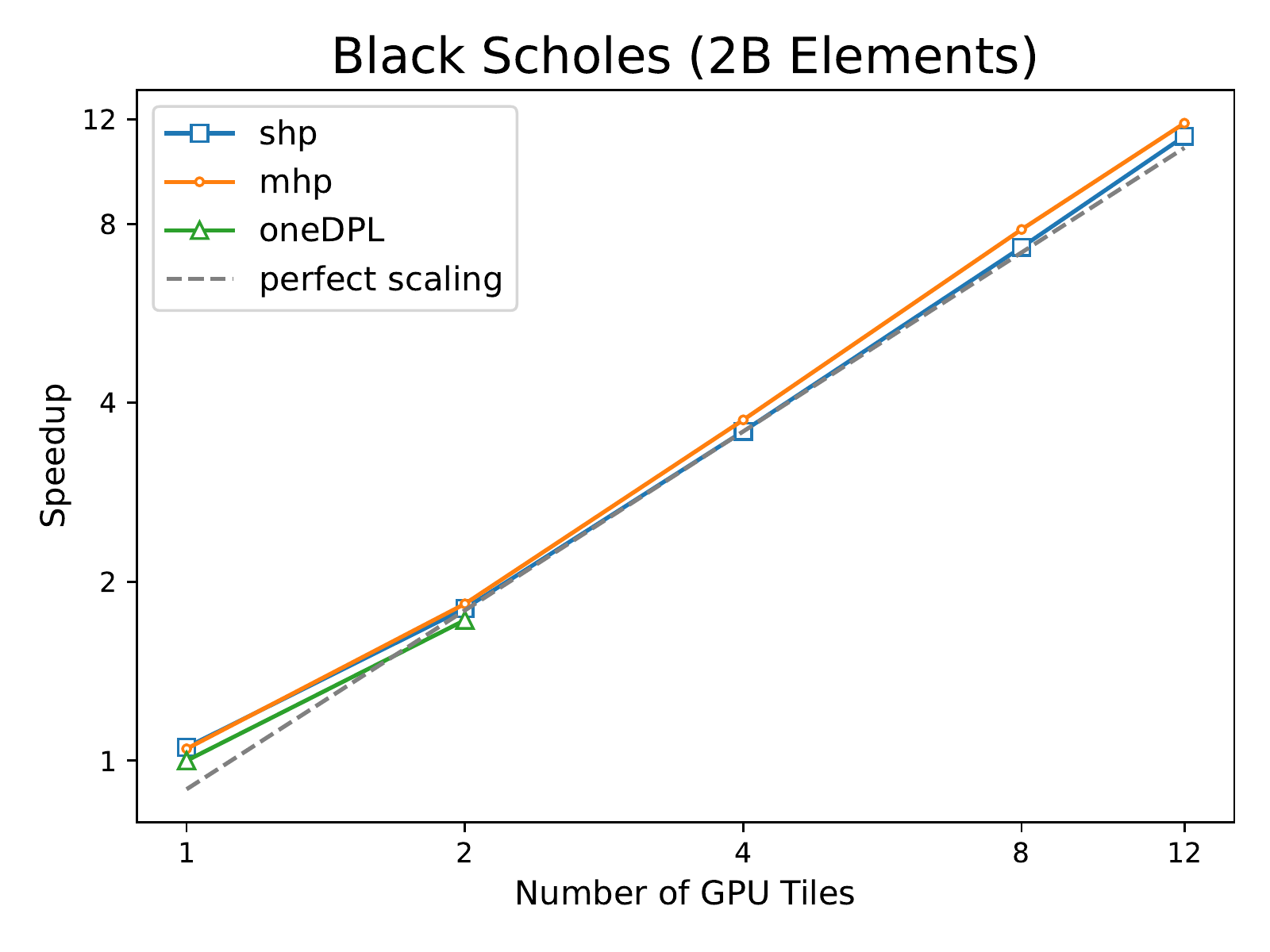}
  \includegraphics[width=0.32\textwidth]{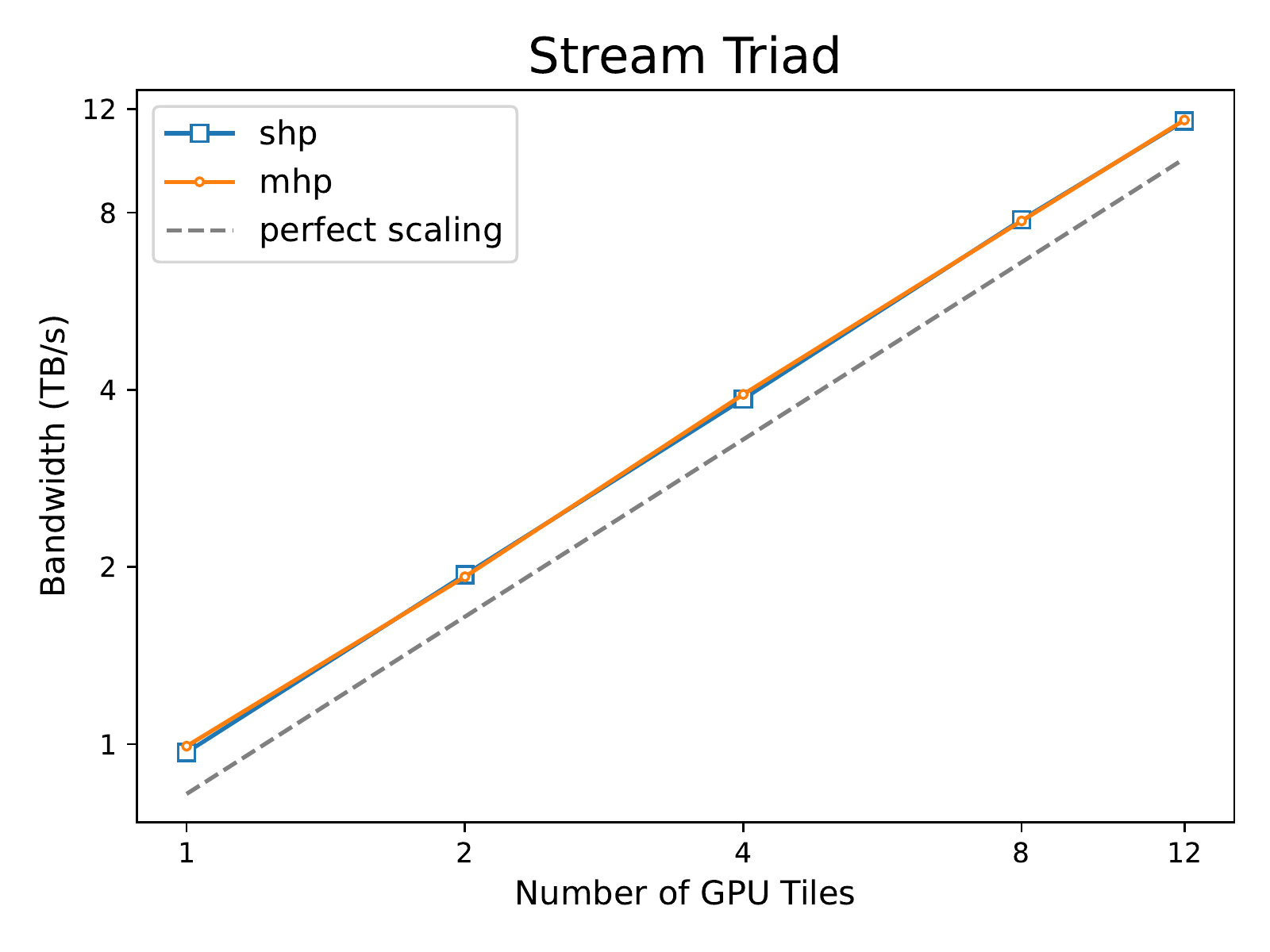}
  \includegraphics[width=0.32\textwidth]{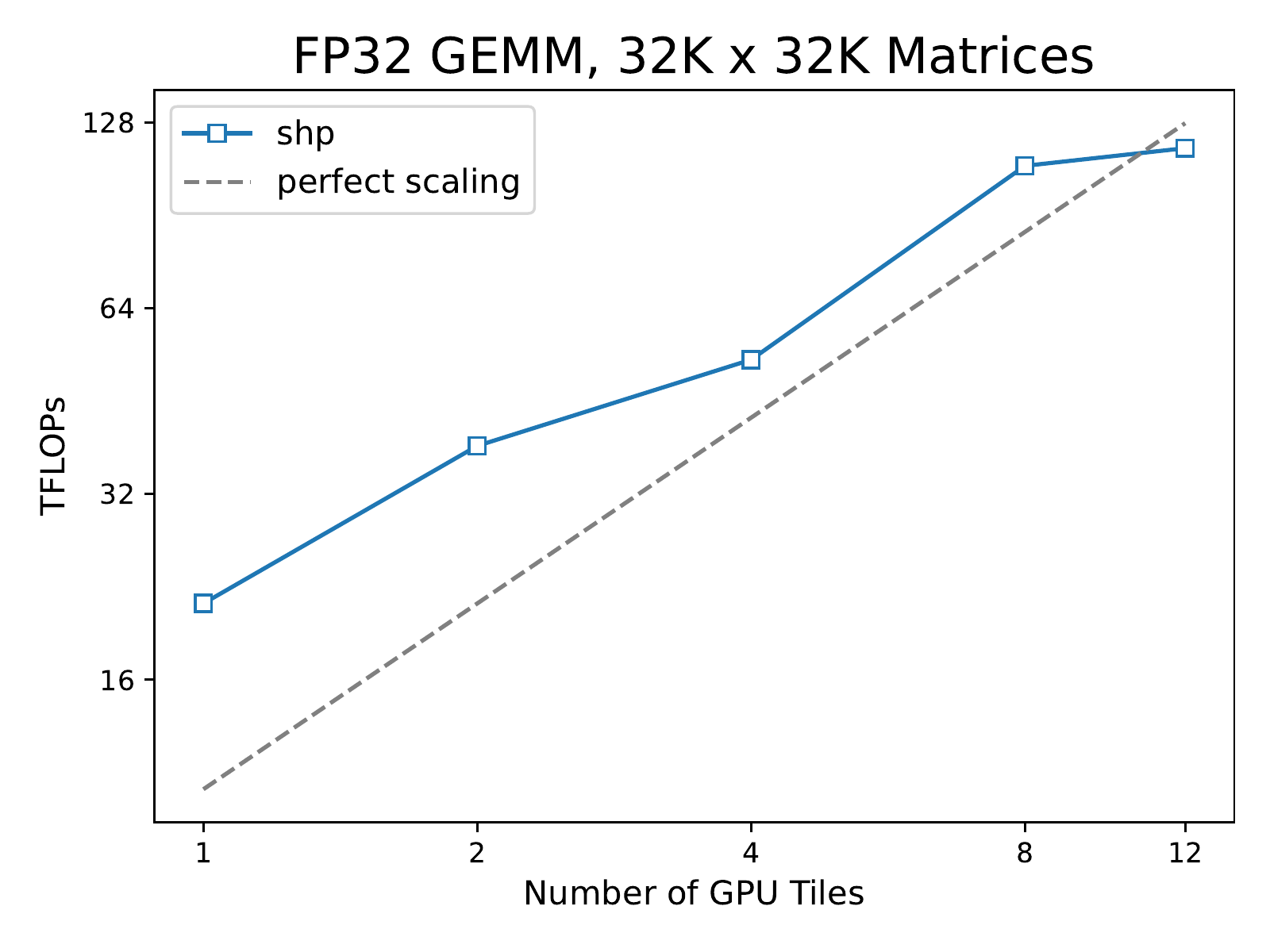}
  \vspace{-0.25em}
  \caption{Distributed ranges benchmarks.  All speedup numbers show strong scaling
           with respect to single tile oneDPL.}
  \vspace{-0.25em}
  \label{fig:dot_product_shp}
  \label{fig:black_scholes_shp}
  \label{fig:stream_shp}
  \label{fig:dot_product_mhp}
  \label{fig:black_scholes_mhp}
  \label{fig:stream_mhp}
  \label{fig:gemm}
\end{figure*}

We tested five algorithms implemented using standard C++ that can be executed
using distributed ranges.  For our baseline, we compare against oneDPL, which
provides vendor-optimized, single-GPU standard algorithms, and oneMKL where
available.
Speedups shown in Figure~\ref{fig:dot_product_shp} are all with respect to
oneDPL running on a single tile.  The bandwidth achieved by the oneDPL baseline
is shown in Table~\ref{table:benchmark_bw}.  The ``perfect scaling'' line plots
a unit slope, offset slightly to avoid crowding the other plotted lines.

\begin{table}[t]
    \centering
    \caption{Performance of single-tile oneDPL baseline.  Single-tile STREAM Copy bandwidth measured at 1063 GB/s.}
    \resizebox{\columnwidth}{!}
    {
        \begin{tabular}{l r r r r r}
        \toprule
        Benchmark & Bandwidth (GB/s) & Percent of STREAM\\
        \toprule
        Dot Product & 903 & 85\%\\
        Reduce & 1025 & 96\%\\
        Inclusive Scan & 806 & 76\%\\
        Black Scholes & 803 & 76\%\\
        \bottomrule
        \end{tabular}
    }
  \label{table:benchmark_bw}
  \vspace{-0.5em}
\end{table}

\paragrapht{Dot Product}
We implement a dot product as shown in Algorithm~\ref{alg:dot_product}.
As discussed in Figure~\ref{fig:dot_product_diagram}, we use a zip view to
zip together corresponding elements of the two vectors into a tuple, followed
by a transform view to multiply the two elements together. Finally, the resulting
view is fed into a \code{reduce} algorithm to sum together all the intermediate results.
As shown in Figure~\ref{fig:dot_product_shp}, dot product achieves performance
competitive with both a oneDPL implementation as well as with the
highly optimized oneMKL library.  While both oneDPL and oneMKL are limited to a
single device, they can be launched on one or two tiles.  This is enabled by the
oneAPI runtime, which can expose a GPU's two tiles as either separate devices or
as a single composite device.  The two tiles have separate HBM2e memories,
leading to a loss in efficiency when naively launching across both tiles
without intelligently partitioning data and work.  oneMKL actually experiences
a drop in performance with two tiles due to frequent cross-tile memory traffic
caused by its temporary accumulation buffer.  As shown, distributed ranges
frequently achieves a small speedup on a single GPU (two tiles) by partitioning
data and work across the tiles.  \shp{} experiences a slight drop in performance
at 12 tiles due to scalability issues in its reduce, discussed momentarily.

\paragrapht{Reduce}
The ``Reduce'' plot in Figure~\ref{fig:dot_product_shp} illustrates the loss in
scalability observed in \shp{}'s reduce algorithm.  Since \shp{} relies on a
single process to launch work, as scale increases, more and more work must be
launched by a single process.  This introduces overhead linear in the number of
tiles, while in the strong scaling problem the amount of work in each kernel
decreases, further exposing any overhead.  For small problem sizes
such as running a single reduce across 12 tiles (5ms runtime), strong scaling
efficiency is reduced.  \mhp{}, which uses a different process for each tile,
sees no such drop in performance.  This drop in efficiency is fundamental
to the fan-out required when launching work from a single process execution model,
and is a motivating reason for users to port their programs to a multi-process
SPMD execution model if they are willing.

\paragrapht{Inclusive Scan}
After an initial drop when moving from one to two tiles, ``Inclusive Scan,''
as shown in Figure~\ref{fig:dot_product_shp}, achieves near linear scaling.
The initial drop is due to increased memory traffic---after the initial
inclusive scan, another pass through the data is necessary to apply the appropriate
offset to each segment.  This plot also demonstrates the importance of distribution
awareness, as oneDPL has significantly reduced scalability on two tiles due to
the increased memory traffic and synchronization required by naively launching
work across two tiles.

\paragrapht{Black Scholes}
Our Black Scholes benchmark solves a partial differential
equation in order to compute the optional prices for European-style options.
The benchmark zips together various arrays
associated with options prices such as the price of an asset and its drift rate,
then performs a series of element-wise operations on these arrays using the
\code{for_each} algorithm.  Similar to dot product, Black Scholes matches
the performance of oneDPL and achieves close to perfect strong scaling.

\paragrapht{Stream}
The stream benchmark performs a series of element-wise updates using the 
\code{for_each} algorithm.  Distributed ranges achieves near perfect scaling.

\subsection{Linear Algebra Benchmark}
To evaluate the performance of our distributed matrix data structure, we ran
strong scaling experiments evaluating the performance of dense matrix multiply
(GEMM) using 32K x 32K matrices.  In our distributed GEMM algorithm, the
\code{get_tile} method is used to retrieve local copies of the tiles necessary
for each local matrix multiplication, which is performed using MKL.
As shown in Figure~\ref{fig:gemm}, our implementation
achieves close to linear strong scaling and a high percentage of peak flops.


\section{Related Work}
The C++ standard library provides provides a collection of standard algorithms
that operate over ranges of data using iterators.  The power of this algorithm
design is that an algorithm can be implemented once generically and then used
with a variety of data structures, so long as their iterators fulfill the
generic iterator requirements.  Algorithms can be optimized for specific
iterator types such as random access, bidirectional, or forward iterators, and
implementations are even free to provide specializations for particular data
structures.  C++17 added parallel versions of some algorithms, allowing
implementations to execute parallel algorithms using multi-threaded and
vectorized execution via standard execution policies.  Implementations can
also provide their own implementation-specific execution policies, which have
implementation-defined behavior, such as executing an algorithm on a particular
execution resource such as a GPU.  Many vendors have provided parallel versions
of these standard algorithms for multi-threaded and single GPU execution,
including Nvidia's Thrust~\cite{bell2012thrust} and CUB~\cite{merrill2015cub},
AMD's rocThrust~\cite{rocthrust2023}, and Intel's oneDPL~\cite{onedpl2023}.

The \code{std::execution} proposal P2300~\cite{p2300} adds a set of standard
facilities for asynchrony and parallelism to the C++ standard library.  These
include schedulers, which can be used to launch work in a particular execution
context, as well as senders and receivers, which can be used to handle
asynchronous events from different libraries.  If \code{std::execution} is
adopted, we anticipate generalizing the \code{rank} customization point in
distributed ranges to associate a range with an execution context or scheduler.
This could potentially allow for greater interoperability between libraries by
combining a generic way of launching work (schedulers) with a common data
distribution (the \code{distributed_range}).



DASH~\cite{Fuerlinger:2016:DASH} and BCL~\cite{Brock:2019:BCD:3337821.3337912}
both implement RDMA-based distributed data structures using PGAS
programming models.  While both BCL and DASH
implement some high-level algorithms on top of their data structures, including
some with APIs very similar to those used by the standard library's algorithms,
these are data structure-specific algorithms, and there is no standard mechanism
for accessing distributed data across different data structures.  They also
completely lack lazily evaluated views like those presented in this work.

HPX~\cite{kaiser2020hpx} provides a parallel runtime with support for standard
C++ asynchronous execution, including support for algorithms and data structures.
HPX's runtime system uses a task-based programming model built on top of an
AGAS memory model, in which objects may migrate from one \domain{} to
another~\cite{kaiser2014hpx}.  HPX provides two containers, a distributed vector
and a distributed hash table, along with a comprehensive set of standard
algorithms that operate on these containers.
Unlike in the distributed ranges model, HPX's
distributed data structures do not directly expose their segments~\cite{hpx2023containers}.
Instead, users can only access a distributed data structure's segments
using segmented iterators~\cite{austern2000segmented}.  With segmented iterators,
a user can transform a normal iterator into a segmented iterator, which
points to a segment, as well as a local iterator, which points to an element
within the segment.  Segmented iterators can be awkward to use and are difficult
to build on top of recursively, since iterators are invalidated once the
range they belong to is destroyed.  HPX does not currently have any support for
views.

SHAD~\cite{DBLP:conf/ccgrid/CastellanaM18} also implements distributed data
structures and algorithms using a task-based programming model and multiple
backends, including Intel TBB, SHAD's multi-node GMT runtime, and HPX~\cite{wu2022shadhpx}.
Similar to HPX, SHAD's data structures expose a segmented iterator design.
SHAD provides a large number of distributed standard library algorithms as well
as domain-specific algorithms for graph computation.  SHAD lacks views, however,
and these algorithms all operate directly on data structures.

STAPL~\cite{Tanase:2011:SPC:1941553.1941586} provides a set of distributed data
structures (pContainers) that can be used with parallel algorithms
(pAlgorithms).  STAPL pContainers expose an iterator-based interface
called a pView.  pViews can represent different ways of iterating
through a data structure.  The C++ language has evolved significantly since
STAPL's pViews were developed using C++03, and elements of pViews'
design limit their performance and usability compared to distributed ranges.
In particular, pViews make heavy use of inheritance, with each pView inheriting
from the base pView class \code{core_view}.  A pView that builds on top of a
base view will store a pointer to the base view, meaning that any operations
on the base view will incur the overhead of a virtual function call as well as
prevent inlining.  This means that in order to achieve good performance,
algorithms must be specifically optimized for combinations of views, limiting
the usefulness of arbitrarily nested views.  In addition, the pView design
requires implementing all views from scratch, removing the possibility of
implementing a single customization point for a pre-existing view, as is done
with many views in distributed ranges.  STAPL provides a number of
custom, domain-specific views for matrix and graph data structures that do not
exist in the C++ standard but are ripe material for future work.

In Python, libraries like Legate NumPy~\cite{10.1145/3295500.3356175} and
Legate Sparse~\cite{10.1145/3581784.3607033} utilize Legion's logical regions data
model and task-based runtime~\cite{bauer2012legion}.  These libraries' deferred
execution model allows them to expose more parallelism, waiting to synchronize
and perform dependency analysis only when data is accessed directly.  In contrast,
the primarily static techniques discussed in this paper are more limited in
their ability to uncover dynamic parallelism, but have very minimal runtime
overheads compared to those of the Legion runtime.

\vspace{-0.5em}
\section{Conclusion}
This paper presents distributed ranges model, which extends the standard ranges model
with concepts providing a standard way to expose the distribution
of a data structure or view.  This provides a flexible, high-performance mechanism
to implement interoperable distributed data structures, algorithms, and views.

\bibliographystyle{ACM-Reference-Format}
\bibliography{references}

\scriptsize
\noindent
\newline Optimization Notice: Software and workloads used in
performance tests may have been optimized for performance only on
Intel microprocessors.  Performance tests, such as SYSmark and
MobileMark, are measured using specific computer systems,
components, software, operations and functions.  Any change to any
of those factors may cause the results to vary.  You should
consult other information and performance tests to assist you in
fully evaluating your contemplated purchases, including the
performance of that product when combined with other products.
For more information go to \url{http://www.intel.com/performance}.

\noindent Intel, Xeon, and Intel Xeon Phi are trademarks of Intel Corporation in the U.S. and/or other countries.

\normalsize

\end{document}